# DIRECT DARK MATER SEARCHES

Neil Spooner[*]

*Institute of Underground Science, Department of Physics and Astronomy, University of Sheffield, Hicks Building, Sheffield S3 7RH, UK*

**Abstract**

For many working in particle physics and cosmology successful discovery and characterisation of the new particles that most likely explain the non-baryonic cold dark matter, known to comprise the majority of matter in the Universe, would be the most significant advance in physics for a century. Reviewed here is the current status of direct searches for such particles, in particular the so-called Weakly Interacting Massive Particles (WIMPs), together with a brief overview of the possible future direction of the field extrapolated from recent advances. Current best limits are at or below $10^{-7}$ pb for spin-independent neutralino coupling, sufficient that experiments are already probing SUSY models. However, new detectors with tonne-scale mass and/or capability to correlate signal events to our motion through the Galaxy will likely be needed to determine finally whether WIMPs exist.

[*] *Email address: n.spooner@sheffield.ac.uk*

## 1. Introduction

Building from the earliest measurements of galaxy clusters, through decades of observational cosmology, ground-based observation of the cosmic microwave background (CMBR) and on in recent times to precision cosmology measurements with the CMBR satellite WMAP and observations of distant supernovae [1,2], the remarkable conclusion we have is that the Universe is geometrically flat ($\Omega \sim 1 \pm 0.04$) but contains only ~4 % ordinary baryonic matter. Even of this 4% only about $1/10^{th}$ is actually visible to us, as stars mainly, the rest is likely composed of cold gas, sub-solar mass "dead" stars and other forms of non-luminous baryonic matter. The great majority of the Universe then is found to be a mix of mysterious dark energy ($\Omega_\Lambda \sim 73\%$), the nature of which is unknown, and non-luminous, non-baryonic dark matter ($\Omega \sim 23\%$) [3,4].

What form this dark matter takes is also so far unknown. However, a generic class of relic particles produced thermally in the early Universe and termed Weakly Interacting Massive Particles (WIMPs), has emerged as a leading possibility [5]. Such, non-relativistic particles would constitute a Cold Dark Matter population that appears required to explain galaxy formation. The observed density required of them in the galaxy, ($\Omega \sim 0.1$), is consistent with the freeze-out relic density appropriate if the mass and cross-section of the particles is determined by the weak scale [6,7]. The CDM model itself provides significant motivation to search for such weakly interacting neutral particles, the appropriate mass range being several GeV to ~TeV. However, two leading theories in particle physics phenomenology greatly enhance this motivation by also, independently, predicting new particles with these features. Firstly, in supersymmetric (SUSY) extensions of the standard model [7], the lightest SUSY particle (LSP), stable in models where R-parity is conserved, provides the required population, known as neutralinos. Theories of universal extra dimensions in which Kaluza-Klein parity is conserved provides a second possible class known as the lightest Kaluza-Klein particles (LKP) [8,9,10].

As an alternative to WIMPs, axions form a further potential candidate, motivated here by extensions to the Standard Model through Peccei-Quinn symmetry as a solution to the strong CP problem [11]. The PQ symmetry is spontaneously broken at a scale $f_a$, with the axion as the associated pseudo-Goldstone boson produced in the early Universe [12,13,14,15]. Though outside the scope of this review we note the rapid progress being made by the ADMX axion search. This experiment is currently setting stringent limits, excluding at >90% c.l. the KSVZ halo axion mass of 1.9 – 3.3 eV and that the local axion dark matter halo mass density is greater than 0.45 GeV/cm3 for KSVZ DFSZ axions [16,17]. For a full review on axions see for instance [18].

In this rich context, research aimed at an explanation of non-baryonic dark matter encompasses a huge worldwide effort. This includes: searches for SUSY at accelerator experiments, such as the upcoming ATLAS/CMS at the Large Hadron Collider and elsewhere; indirect searches through attempts to detect the products of neutralino self-annihilation in astrophysical objects, such as neutrinos from the Galactic centre, the Sun and Earth; and direct particle searches for both axions and WIMPs using experiments in the laboratory. Focus will be placed here on the latter case - arguably the best motivated candidate studied by the best generic technique. That is, efforts toward detection of relic WIMPS in the galaxy via their direct interactions in detector target materials on Earth [19,20]. For a recent review of indirect searches for neutralinos see for instance [21] and for accelerators see for example [22,23]. The firmest indirect neutralino limits from high energy neutrinos coming from the Sun are currently set by Super-Kamiokande [24].

## 2. Requirements for direct detection

For direct detection our starting point is that the Galaxy contains a halo of WIMPs normally assumed to be of spherical isothermal form with local density 0.3 GeV $c^{-2}$ $cm^{-3}$, an escape velocity of 650 km $s^{-1}$, with *rms* velocity 279 km $s^{-1}$ and relative Halo-Earth velocity of 235 km $s^{-1}$ [20]. The basis for detection is then elastic scattering of these neutral, non-relativistic particles, off target nuclei in a suitable detection medium, such that the energy transferred as the resulting nuclear recoil passes through the material can be observed, usually as either ionisation, scintillation or heat (phonons). Kinematics and the likely mass range and velocity of the particles implies a nuclear recoil spectrum with energy below ~100 keV, with exponential form rising to low energies and with no spectral features. This characteristic, together with the expected low interaction rate of likely 1-$10^{-6}$ event $kg^{-1}d^{-1}$, dictates three core requirements of WIMP detector technology: low energy threshold (<10 keV$_{recoil}$); potential for target masses of >10 kg; and low particle background of all types. The latter implies the need for a deep underground site to reduce cosmic ray muon-induced neutrons, that could otherwise produce nuclear recoils indistinguishable from WIMPs; use of additional passive and active gamma and neutron shielding; and detector construction using materials with greatly reduced radioactive U, Th and K content.

The coupling of these non-relativistic WIMPs has two terms, a scalar, spin-independent (SI) part and an axial spin-dependent (SD) part [25]. For most SUSY models SI provides the dominant coupling and hence highest rate. This is because although neutralino-nucleon cross sections are mainly much smaller for the SI case [26], coherence across the nucleus results in constructive interference which greatly enhances the WIMP-nucleus elastic cross section for high A targets. The opposite is true for SD where the axial coupling to nucleons with opposite spins interferes destructively. Effectively, sensitivity to SD interactions thus requires a target isotope with an unpaired nucleon, either proton or neutron. Although generally lower sensitivity is implied for the SD case this is not true for all neutralino models. SD targets are certainly required if the full WIMP parameter space is to be studied and the widest investigation of any signals undertaken [27,28,29].

The stringent requirement for low background, bearing in mind, for instance, that typical ambient environmental gamma fluxes can produce event rates $>10^5$ times higher than the expected WIMP signal rate in an unprotected detector, has focussed world attention on technologies that can actively reject electron recoil events, whilst maintaining high sensitivity to nuclear recoils. This is possible in principle because the latter have typically x10 higher dE/dx values [20]. In practice, few technologies can make use of this physics, the prime ones being: (1) low temperature ionisation/phonon or scintillation/phonon detectors in which the ratio of event-produced ionisation or scintillation to phonons is measured in suitable cryogenic materials such as Ge or Si (ionisation) and $CaWO_4$ (scintillation); (2) noble liquid gases, notably xenon and argon, in which scintillation and ionisation is measured simultaneously. A moderate level of discrimination can also be achieved in specific scintillators such as NaI(Tl), CsI(Tl), liquid Ar and liquid Xe (see sec. 3).

Although recoil discrimination, and background reduction, appears feasible in such technologies there remains the issue, given the lack of spectral features in the recoil spectrum, of how to determine in a clear way whether any remaining counts are due to WIMPs from the galaxy and not either nuclear recoils from an unaccounted for background (such as neutrons or surface interactions) or a detector artefact. There are two prime possibilities for addressing this using galactic dynamics. Firstly, at least for the standard halo model, the Earth's motion through the Galaxy implies an expected seasonal modulation in the recoil spectrum (flux and shape) [30,31]. This is because the component of the Earth's solar orbital velocity in the direction of our galactic motion (orbital plane inclined at 60º) either adds to or subtracts from the galactic orbital velocity depending on the season. Secondly, thanks also to our galactic orbital motion (~235 kms$^{-1}$), we would expect the direction of the WIMP-induced nuclear recoil tracks themselves within a target to be dominantly opposite to our direction of motion (in galactic coordinates) [32,33] (see sec. 8). Information may also be gleaned by comparing different targets, since WIMPs interact differently with target nuclei of different A [20] (see sec. 8), different technologies and different sites.

Unfortunately, the annual modulation effect is very small, typically a few %, requiring already at least ton-scale detectors to obtain sufficient event statistics for a viable search [20]. Such small effects are also at risk of being masked by detector characteristics always vulnerable to natural seasonal changes in the environment. The recoil direction effect is far more powerful in principle. Only of order a few 10s of WIMP events are required in such a recoil direction sensitive detector to identify them as of galactic origin (see sec. 8). Furthermore, the angular distribution of WIMP-induced recoil tracks can not be mimicked by any terrestrial backgrounds since we would expect a sidereal (not daily) modulation of the WIMP-induced track directions and an average "washed-out" isotropic distribution of any background in the galactic frame. The challenge here however is the likely need to use low pressure gas detectors that would then need to be very large in volume (100s m$^3$).

Built on these basics a wide variety of experiments have and are being run worldwide. Fig. 1 provides a summary of results from recent key examples, given here as an exclusion plot of WIMP-nucleon cross section vs. WIMP mass for the SI case, assuming the standard halo model as above. Referring to this the following sections outline the current status and possible future scenarios. Note this is necessarily selective and likely tinged by personal bias - for a wider view we refer to recent workshop proceedings such as [35]. Note also that some of these results are yet to be published. They are shown here for the sake of completeness as reported in workshops and preprints, leaving discussion of their validity out of the scope of this review.

## 3. Semiconductors

Ionization detectors, in the form of low background germanium (HPGe) and silicon diodes used for double beta decay searches, provided the first limits on WIMP interactions [36,37]. Such experiments were vital to ruling out early candidates for WIMPs including, Cosmions and heavy Dirac neutrinos [38], but as a technology they suffer from an inability to distinguish between gamma background events and the nuclear recoil events of interest. This is partly compensated for by the possibility of high radio-purity in Ge which has allowed more recent experiments such as HDMS and IGEX to set interesting limits (see fig. 1) [39,40]. The IGEX experiment at Canfranc used 2.1 kg of purified Ge with a 20 cm thick Pb gamma shield inside a muon veto. The detector achieved an eventual background of 0.21 KeV$^{-1}$kg$^{-1}$d$^{-1}$ at 4-10 keV [40].

Next generation HPGe detectors aim at further reduction in activity, for instance by possibly x1000 using novel techniques such as crystal growth underground to reduce cosmic-ray spallation activity. There is also prospect for active rejection of Compton scatter events using segmentation to provide position sensitivity and use of active coincidence Compton vetos. Key ideas have been proposed by GEDEON [41], following from IGEX, the GERDA experiment at Gran Sasso [42] and MAJORANA [43]. All these are primarily aimed at neutrinoless double beta decay detection. The GERDA detector incorporates the novel prospect of using direct submersion in liquid argon or nitrogen (an idea tested by GENIUS-TF [44]) and, for argon, a possible active veto.

## 4. Scintillators

Pulse shape analysis (PSA) in certain organic and inorganic scintillators has been known to allow discrimination against low dE/dx events (electron recoils) for many decades [45]. NaI in particular was turned to advantage for WIMP searches by the UKDM collaboration, using cooled undoped NaI and later NaI(Tl), and by BPRS/DAMA [46,47,48]. Unfortunately, the light output (~40 photons per keV$_{electron}$ in NaI(Tl)) is too low for event by event discrimination at low energy even though the quench factors (~9% for I and ~25% for Na) are relatively high [49]. Statistical methods, combined with material purification to reduce intrinsic activity [50], can be used and were successfully implemented in, for instance, NAIAD to produce significant new limits [51]. Nevertheless, the discrimination power with statistical analysis is limited. The DAMA experiment thus later turned to using NaI in simple counting mode as a means of searching instead for an annual modulation signal, with no nuclear recoil identification applied. They found evidence for a modulation, reporting the discovery of WIMPs in 1997 [52].

The final DAMA result from a total of 107,731 kg day accumulated with 9 low background 9.7 kg NaI(Tl) crystals [53] remains the only claimed direct observation of WIMPs, corresponding to a mass of ~52 GeV and cross section ~7.2 × 10$^{-6}$ pb (for standard halo model assumptions). However, the result appears in contradiction with several other experiments including the bolometric Ge experiments of EDELWEISS and CDMS, and the liquid xenon experiment ZEPLIN [54,55,56] (see sec. 5 and 6). This contradiction appears to hold regardless of the halo model or if spin-dependent interactions dominate [57,58,59] though there remains debate as to whether fine tuning of models can allow compatibility, particularly for the spin-dependent case. The DAMA group is now running an expanded array, the 250 kg LIBRA experiment [60,61]. Following closure of NAIAD no direct test is being made of the result with an independent NaI-based detector, although the Zaragosa/Canfranc group is building a 107 kg NaI experiment, ANAIS, to address this gap [62,63] and the KIMS experiment is producing competitive limits with CsI [64].

More recently there has been interest in other inorganic scintillators, notably CsI(Tl), CaF$_2$(Eu) and, for instance, CaWO$_4$. The former, now developed for the KIMS experiment in South Korea [64], has intrinsically better pulse shape discrimination than NaI(Tl) [65] but potentially higher intrinsic background, in particular due to $^{137}$Cs from nuclear fallout. Nevertheless, encouraging results have been obtained by taking care in material selection and purification. CaF$_2$(Eu) has relatively poor discrimination [66]. CaWO$_4$ and similar compound inorganics are not efficient scintillators at room temperature but operate well at mK temperature. CaWO$_4$ has become an integral part of the CRESST bolometric experiments in which scintillation light is measured simultaneously with heat [67] (see sec. 5). For recent measurements of the quench factors here see ref [68].

Certain organic crystal scintillators such as stilbene, plastics and liquids also demonstrate pulse shape discrimination [45]. They have the advantage of potentially relatively low cost per kg and high radio-purity. However, their composition is dominated by H, C and possibly F or other light elements. This results in poor quench factors, typically 2% or less [69,70], and poor kinematic coupling to WIMPs leading to very poor sensitivity relative to the inorganics like NaI(Tl). Nevertheless, there has been significant interest in the organic crystals, in particular because some of these, for instance stilbene and anthracene, yield a response that is dependent on the direction of the contained recoiling nucleus, at least as measured using alpha particles. This yields a rare example of a technology relevant to the possibility of a direction sensitive WIMP detector [71,72] (see sec. 8).

Whilst the lack of powerful recoil discrimination is a disadvantage for experiments based purely on scintillation detection there is a potential advantage for spin-dependent sensitivity due to the greater possibility of using spin nuclei, particularly iodine (e.g. in NaI, CsI) and fluorine (e.g. in $CaF_2$). This has allowed NAIAD to maintain competitive limits for WIMP-proton coupling [73]. Finally, a remaining class of scintillators of interest is the liquid noble gases, notably liquid xenon and argon. These are covered in sec. 6.

## 5. Bolometers

At low temperature the heat capacity of a dielectric crystal goes as $T^3$. Thus at mK temperatures the small energy deposition from a nuclear recoil can yield a measurable proportional increase in crystal temperature [20]. Some of the earliest techniques investigated for WIMP dark matter detection were based on this, where energy released by particle interactions can be observed as phonons or quanta of lattice vibrations. Work started on this idea in the 1980's (see for instance [74]), the original motivation being in part the prospect of obtaining very low recoil energy thresholds and high energy resolution, due to the meV level of quantisation involved [75]. However, it was soon demonstrated, first in Si [76] and then in Ge [77], that phonon detection could be combined with simultaneous detection of ionisation to provide also a powerful means of discrimination against electron recoils, on an event by event basis. This arises because the proportion of energy observed in the two channels is dependent on the event dE/dx - a high dE/dx event, such as a recoiling nucleus, produces proportionally more heat than ionisation (the ionisation is quenched). For instance, the ratio of ionization to recoil energy (the ionisation yield) for Ge recoils in in Ge is ~0.3 of the value for electron recoils above 20 keV [54].

Whilst bolometers without collection of ionisation have proven quite useful for dark matter searches, the hybrid technique of simultaneous ionisation and phonon collection with its capability for background rejection has been pushed harder. Most notable is the CDMS collaboration (at Soudan mine) and EDELWEISS-I (at Frejus) [54,78,79,80] (see Fig. 1). The latter used 320g Ge crystals operated at 17 mK with NTD-Ge thermometric sensors attached for the heat signal and Al electrode used to collect the charge. 10 cm of Cu and 15 cm of Pb where used to shield the cryostat from rock gamma-ray background with an additional 7 cm Pb inside and a total of 30 cm paraffin outside the entire setup to reduce rock neutrons. A variety of detectors were tried in EDELWEISS-I with several runs completed from 2000 until Mar. 2004. These yielded a total exposure of 62 kg days, the main results coming from three crystals with recoil energy threshold of 13 keV or better over 4 months of stable operation. Fig. 1 shows the limits produced. After cuts a total of 40 nuclear recoil candidates were recorded in the range 15-200 keV with 3 events between 30 and 100 keV, most likely due to remaining background neutrons or surface electrons.

The CDMS experiment operates towers of Ge and Si crystals each 1 cm thick and respectively of mass 250 g and 100 g. These are mounted in a dilution fridge and shielded mainly by 22.5 cm of external Pb and 50 cm of polyethylene. A 5 cm layer of plastic scintillator is used to veto any events coincident with cosmic muons (necessary here due to the relative shallowness of the Soudan site at 2080 m.w.e.). Charge electrodes are used for ionisation collection as in EDELWEISS but here athermal phonons are detected using superconducting transition edge sensors, applied by photolithography to the crystal surfaces. This design has the potential advantage of providing depth position sensitivity, via measurement of the phonon pulse risetime, and hence the possibility of rejecting surface electron events that could otherwise contaminate the signal region, as suspected in EDELWEISS-I. Two towers were operated in 2004 yielding an exposure for 10-100 keV$_{recoil}$ of 34 kgd Ge and 12 kgd Si. No events were observed in the Si and only one event, consistent with the expected surface event background, in the Ge, yielding a 90% c.l. spin-independent upper limit in Ge of $1.6 \times 10^{-7}$ pb at 60 GeV c$^{-2}$ WIMP mass (see Fig. 1).

As an alternative ROSEBUD [81] and CRESST [67,82] have developed detectors in which scintillation light is measured in coincidence with heat, in particular using CaWO$_4$ [82]. Here a silicon wafer of 30 x 30 x 0.4 mm with tungsten thermometer is used to detect the photons and a 8 x 8 mm, 200 nm thick superconducting evaporated film used as the heat sensor. Although only 1% or less of the energy deposited is detected as photons this is much higher than feasible at room temperature and is sufficient to produce an energy resolution comparable to NaI(Tl). Results so far have been obtained with two 300g crystals at the Gran Sasso Underground laboratory with a total exposure of 20.5 kg days. This revealed 16 events in the range 12-40 keV consistent with the expected background from neutrons given that the experiment did not have a neutron shield. The resulting limit, due to W recoils (see Fig. 1), is comparable to others in the field including EDELWEISS.

Notable in the pure cryogenic detector field is the work of the Milan group through the CUORE/COURICINO experiment at Gran Sasso [83,84]. CUORE is designed primarily for neutrinoless double beta decay searches. CUORE demonstrates one particular advantage of pure cryogenic detectors over the hybrid types. The latter is essentially restricted to Ge and Si because only these are found to have sufficiently high electron-hole transport at mK temperatures. The non-hybrid technique, at the expense of throwing away recoil discrimination, is open to a much greater variety of target crystals, for instance TeO$_2$ for CUORE, LiF, Sapphire and others have been demonstrated. CUORE aims to build an array of 988 TeO$_2$ cryogenic crystals with total mass ~750 kg, building on the first stage CUORICINO experiment already operated with 62 (~40.7 kg) crystals. Although CUORE will have exceptionally high target mass, competitive WIMP limits will only come through more work to suppress intrinsic crystal backgrounds [85].

All these cryogenic experiments are now progressing towards significant upgrades. CDMS is proposing 25 kg and a possible move to the deeper SNOLAB site. EDELWEISS is progressing towards a more ambitious phase II with up to 120 detectors and CRESST is upgrading to allow 33 CaWO$_4$ detectors totalling 10 kg. However, as outlined in sec. 7, it is likely that even greater target mass will be needed, possibly at the tonne-scale or larger.

## 6. Liquid noble gases

Whilst a large world effort has been devoted to cryogenic bolometers over many years, linked now to quite an industry in alternative applications, there has been recent rapid growth in liquid noble gas technology for WIMP searches. Most notable has been liquid xenon (LXe), started by DAMA/Xe [86,87], but also recently liquid neon [88] and, in particular, liquid argon. A prime motivation here has been improved low background discrimination combined with tonne-scale target mass at reasonable cost (see sec. 7). LXe has particularly good intrinsic properties for WIMP detection including: high mass (Z=54, A=131.3) - yielding a good kinematic match to likely WIMP candidates; high scintillation and ionisation efficiency (~46 photons/keV at 178 nm); and high radiopurity, enhanced further by the availability of liquid gas purification techniques. However, of greater importance is the recoil discrimination achievable. This is possible firstly, as in NaI(Tl), by simple pulse shape analysis (PSA) of the scintillation light. This is the basis for the single phase LXe XMASS experiments in Japan [89] and of ZEPLIN I [56]. The latter detector, comprising 3.2 kg of active LXe viewed by 3 PMTs, accumulated 293 kg days during operation at Boulby mine until 2002, producing significant limits with this technology (see Fig. 1).

The XMASS group have also run a 100 kg prototype PSA detector and are aiming to achieve higher sensitivity by constructing an 800 kg experiment for operation in Kamioka mine [89]. However, more powerful discrimination is feasible in LXe by recording also the ionisation produced and hence the ionisation/scintillation ratio. This arises because for nuclear recoils the ionisation signal (termed S2) is quenched significantly more than the primary scintillation (S1) relative to electron recoils of the same energy. This is being implemented by the ZEPLIN II/III, XENON 10/100 and XMASS II (two-phase) experiments, all aiming to achieve higher sensitivity with lower fiducial mass than likely required with PSA alone [90,91,92,93,94]. Collection of event ionisation can in principle be achieved in the liquid phase of xenon (see sec.7), but obtaining stable operation with the high avalanche fields required (~1 MV $cm^{-1}$) is a challenge [95]. The current generation therefore uses two-phase operation, the charge first drifted out of the liquid into a gas phase amplification region, used to produce an electroluminescence observable by PMTs as the second (S2) light pulse.

Exciting progress has been made recently with this two-phase LXe technology with both ZEPLIN II (Boulby) and XENON 10 (LNGS) announcing new leading limits (see Fig. 1), yet to be published. As the largest two-phase detector so far the 30 kg ZEPLIN II successfully accumulated over 1 tonne.day of data during 2006 before all cuts. This detector, using a relatively simple design comprising a PTFE lined LXe chamber viewed by 7 PMTs, has pioneered operation of bulk two-phase xenon. XENON10, with 15 kg active volume, has a more complex design involving PMTs viewing both top and bottom (48 and 41 Hamamatsu R8520s respectively). This detector achieved light collection >2 p.e./keV and excellent stability over 9 months in 2006/7, accumulating 136 kg.d of data after cuts. This recently allowed new world best limits to be derived, set provisionally at 5.5 x $10^{-8}$ pb at 100 GeV $c^{-2}$ assuming background subtraction [93]. Figure 2 gives a plot of ionisation yield vs. energy from XENON10 showing the gamma background region around logS2/S1=2.5 and the signal region below used to set the current limits.

Despite the current interest in LXe the intrinsic discrimination may not match that of bolometers. In this respect liquid argon (LAr) may provide better prospects. Some properties of LAr are inferior to LXe for WIMP searches, notably the lower Z, A (18, 40) and the need to use wavelength shifter for the VUV scintillation light (~40 photons/keV at 135 nm). However, both pulse shape discrimination and two phase primary/secondary discrimination are now known to be more powerful, capable in principle of combined discrimination factors up to ~$10^8$ [96,97]. LAr is also a factor ~x400 lower in cost. Based on this the WARP collaboration has built and deployed a 3.2 kg LAr experiment at LNGS and recently reported a sensitivity near $10^{-6}$ pb at 100 GeV $c^{-2}$ for an exposure of 96.5 kg days [98]. WARP is currently constructing a larger 140 kg detector with a full active Compton veto. Other LAr experiments are also under construction including ArDM, CLEAN and DEAP [99,100,101,102]. ArDM involves two-phase operation but with ionisation readout using direct collection by LEMs (Large Electron Multipliers) in the gas phase. The other designs are based on single phase PSA plus self-shielding, akin to the XMASS concept with liquid xenon.

Although ZEPLIN I, II, XENON 10 and WARP have shown excellent progress, significant issues remain with liquid noble gases. Firstly, the quench factor for nuclear recoils remains poorly determined. Measurements for LXe in zero field have indicated 0.13-0.23 (10-56 keV) [103] but higher values have been claimed [104]. For argon there are several conflicting results, for instance [98,101]. Secondly, LXe has not yet demonstrated recoil discrimination competitive with cryogenic technology. Populations of events are observed to spread from the gamma region into the signal region. This may reflect the youthfulness of current detector designs but may be intrinsic. It is possible that spontaneous single electron emission occurs in the liquid, producing secondary electroluminescence with minimal ionisation signal and yielding events indistinguishable from nuclear recoils. The ZEPLIN III detector, currently being installed at the Boulby, has improved light collection and higher drift fields than ZEPLIN II and will be used in part to investigate this prospect [92].

The background issue above may also be present in LAr. However, for argon there is a more important issue to resolve, the presence of radioactive $^{39}$Ar [105]. Produced by cosmic ray spallation in the atmosphere, this yields in industrial argon a beta background of ~1 bq/lt. Discrimination of $10^{10}$ would be sufficient in principle to cope but then data acquisition deadtimes in a ton-scale detector would likely be unmanageable. Calculations show that argon from deep gas wells, shielded from cosmic rays, could provide an economic source of so called dead-argon. Activation times on the surface are long enough that once brought to the surface there is sufficient time to construct and deploy an experiment [106].

7. Tonne-scale concepts and alternative techniques

After over two decades of development, WIMP experiments with target masses of kg-scale are reaching sensitivities improved by about 4 orders of magnitude, probing well into SUSY favoured parameter space. This achievement has been accompanied by a continuing rise in the number of experimental scientists involved, now >300. There has been an expansion of interest in new and emerging technologies, not just liquid noble gases but others not detailed here, including superheated droplet detectors

(SSD), specifically SIMPLE and PICASSO, and the MACHe3 detector that uses superheated He [107,108,109,110,111]. The SSD experiments, though use of Fluorine-loaded targets, show particular promise for spin-dependent sensitivity and are producing interesting limits. However, whilst all this activity together reflects substantial maturity, a crossroads has likely been reached in the field.

Firstly, it is pretty certain, setting aside claims by DAMA, that favoured spin-independent coupled dark matter does not exist with cross-sections $>\sim 2 \times 10^{-7}$pb (see Fig. 1). Meanwhile, theoretical predictions for both neutralinos and LKPs reach $<10^{-11}$pb [112,113,114,8]. Thus next generation experiments must not only achieve further background suppression but also be capable of tonne/multi-tonne masses, simply to ensure a statistically observable signal rate. This represents a major leap, implying significantly higher costs and likely larger collaborations. Secondly, for such large detectors it can be argued that though active gamma discrimination remains important, greater emphasis is needed on material purification, passive shielding of external backgrounds and on searches for additional features in the data to show that remaining events are non-terrestrial signals and not, in particular, neutrons.

The latter argument arises as follows: assuming next experiments are at sufficient depth to avoid muon-induced neutrons then gammas and neutrons from U/Th chains in the environment and detector will dominate background. For the relevant energy range, <200 keV, such contamination produces typically $10^5 - 10^6$ more gammas than neutron induced nuclear recoils [20]. The levels of detector sensitivity required for tonne-scale experiments now imply that gamma backgrounds must be comparable with or lower than the neutron rates, such as could be achieved by neutron/gamma discrimination of $10^5 - 10^6$ (the rate for fast neutrons from the rock at Boulby, for instance, has recently been measured to be $1.72 \pm 0.61(stat.) \pm 0.38(syst.)) \cdot 10^{-6}$ cm$^{-2}$ s$^{-1}$ above 0.5 MeV [115]. Thus neutron induced recoils, which can not be distinguished from WIMP interactions, naturally will be the dominant particle background. Detector position sensitivity may help, by allowing rejection of multi-scatter events [116]. However, ultimately reliance will be needed on passive neutron shielding and material purification plus WIMP signal identification via: (1) use of at least two targets/technologies with different A and different systematics; and/or (2) correlation of events with Galactic motion by observation of annual modulation or a directional signal. The former relies on the different behaviour of WIMP and neutron scattering cross section vs. A to deduce that a signal is not neutrons. The latter allows direct identification of events as of extra-terrestrial origin.

Following these notions, similar to arguments adopted in neutrino physics by, for instance, Borexino and SNO [117,118], it is natural to consider larger WIMP detectors with (near) spherical design, a central fiducial zone containing minimal detector components other than the target material, and an integral passive outer shield. This is the basis of the XMASS (Xe), CLEAN (Ar and Ne) and DEEP (Ar) scale-up programmes (see sec. 6). Single phase liquid noble gases are used here with photons recorded by photomultipliers in the outer region, pointing inwards (see for instance Fig. 3). This allows both position information (fiducialisation) and some recoil discrimination via pulse shape analysis (PSA), but the dominant theme is bulk passive shielding.

The reliance on PSA and the presence of PMTs (with relatively high radioactivity) is a potential limitation for these experiments. Replacement of PMTs is a possibility by using an internal photocathode such as CsI to convert photons to electrons for collection by charge readout in the gas phase, for instance using micromegas or GEMs [119]. Current two-phase programmes, with greater discrimination potential, are also being developed for scale-up, for instance WARP [98], LUX [120], XENON100 [121] and ArDM [99], which is already at the ton-scale and uses LEM charge readout. However, use of gas-phase electroluminescence is not well suited to the benefits of a pure spherical concept due to the need for a top gas layer. An alternative hybrid design has been suggested, termed CORE in which the ionisation signal is recorded directly in the liquid phase at a point gain region central to a sphere [122]. Such a spherical TPC concept has in fact already been realised in the gas phase by NOSTOS [123]. Use of high pressure noble gas, as developed by SIGN, may itself provide an alternative class of scale-up technique [124].

Cryogenic technology is not so well suited to the massive self-shielding spherical concepts above. Nevertheless, scale-up to tonne-scale is planned here also, making best use of the high discrimination power demonstrated notably by CDMS, EDELWEISS and CRESST. Two particular efforts are foreseen, SuperCDMS [125] and EURECA (European Underground Rare Event search with Calorimeter Array) [126]. The former will use Ge and Si ionisation/thermal technology like CDMS in a staged expansion from 27 kg to 145 Kg and eventually to 1100 kg by 2015, either at the US DUSEL, if built, or SNOLAB in Canada. EURECA represents a merger of EDELWEISS, CRESST with further new groups to develop a 100-1000 kg array using various targets, possibly both ionisation/thermal and scintillation/thermal ideas. For both experiments a priority will be the need to develop improved detectors, in particular to allow better rejection of surface events, for instance through event position reconstruction or improved analysis, and to reduce unit costs.

## 8. Directional Detectors and Proof of a Galactic Signal

The scale-up programmes, assuming more than one becomes reality, in part address the signal identification issues noted in sec. 7 by opening the way to examining the A-dependence of a potential WIMP signal. However, definitive proof that a signal is of galactic, and not terrestrial, origin can only be achieved by correlating in some way the events with our motion through the Galactic WIMP halo (see sec. 1) [32]. This has been the objective of DAMA/LIBRA by making use of the small predicted annual modulation in flux/energy [60]. However, a much more powerful, though technologically challenging, possibility is to correlate in 3D the physical direction of nuclear recoils in a target with our motion. This is the motivation behind the DRIFT, MIMAC, NEWAGE and other low pressure gas Time Project Chamber (TPC) r&d programmes [127,128,129,130,131]. Calculations show that in principle only a few 10s of WIMP events are needed to prove a Galactic origin. Furthermore, a powerful sidereal day modulation of the signal is expected in the laboratory frame, impossible to be mimicked by any terrestrial background [33,132, 133,134].

Much progress has been made here by the US-UK DRIFT collaboration using negative ion $CS_2$ Time Projection Chambers (TPCs) at Boulby mine. The latest version, DRIFT II, comprises 3 units of 1 m$^3$ of $CS_2$ at 40 Torr (170 g fiducial mass) [128]. The reduced pressure is needed so that nuclear recoil tracks are extended to a few mm, sufficiently for observation by multi-wire proportional counters (MWPC) readout. Negative ion gas is used to minimise track diffusion without the need for expensive magnets, the $CS_2$ (and possible additives) providing also a multi-A target. The detector contains a 1 $m^2$ central high voltage cathode plane and two back to back drift regions of 50 cm depth, each read out by a 1 $m^2$ MWPC comprising planes of 20 µm wires at 2 mm pitch.

DRIFT II demonstrated stable, neutron shielded, long-term running during 2005/6. Operation is by remote control at room temperature, with no cryogenics or complex services. The in-built sensitivity of the TPC technology to particle dE/dx (ionisation charge density) allows exceptional electron track rejection (>$10^5$), sufficient that no gamma shielding is required for DRIFT II. More importantly analysis of event drift time, MWPC anode hits and induced signals on the orthogonal grid planes allows in principle full x,y,z 3D reconstruction of ionisation tracks down to 300 NIPs (number of ionising pairs) or <10 keV$_{recoil}$. Fig 4. shows 3D reconstruction of a typical S-recoil event resulting from a neutron elastic scatter [135].

As currently the only known route to significant recoil direction sensitivity, TPC technology holds exceptional power for WIMP physics and possibly the only route to a definitive galactic signal. However, there are several challenges to address. The requirement for use of low pressure gas implies large volume detectors will eventually be needed, possibly 1000s $m^3$. For this new charge readout technologies such as bulk Micromegas and Gas Electron Multiplier (GEM) planes are under development to reduce spatial resolution and hence allow high pressure, lower volume, operation [136,137]. A further issue is the desirability of achieving track head to tail discrimination. Orientation of a recoil track alone provides significant directional information but a factor ~x10 greater sensitivity can be achieved in principle if the head can be distinguished form the tail [33]. Whilst more careful data analysis is required to demonstrate feasibility it is clear now from recent detailed simulations that this also should be possible [138,139] (see Figure 5). There is clear overall intrinsic head-tail asymmetry and the predicted range agrees well with experimental data. Work is underway to determine whether this asymmetry is observable in a practical detector set-up and a new international cooperation, CYGNUS, has formed to address the design challenge of a very large directional dark matter detector [140].

## 9. Conclusion

In summary, great progress has been made toward detection of particle dark matter in recent years, notably through development of cryogenic detectors but also liquid noble gas experiments now beginning to set the best limits. Much effort has been placed on producing technology with recoil discrimination against gammas. However, it is not necessarily clear that gamma discrimination alone will be sufficient to prove the presence of WIMPs. Proof that any remaining signal is in fact from extra-terrestrial dark matter and not neutrons or some other un-determined terrestrial background or artefact will be vital. Multi-ton detectors with optimal passive shielding to achieve sufficient count rate are being developed but there is also a potential route toward the needed definitive galactic signal through new recoil direction sensitive technology.

## Acknowledgments

The author wishes to E. Daw, V. Kudryavtsev and P. Majewski for many useful discussions. Thanks also to the ILIAS integrated activity (contract R113-CT-2004-506222) workpackage N3-LX, N3-AD, N3-BSNS, the INTAS programme (grant number 04-78-6744) and funding provided by the UK Particle Physics and Astronomy Research Council (PPARC) for financial support.

## References

[1] D.N. Spergel et al., astro-ph/0603449.

[2] S. Perlmutter et al. *Astrophys. J.,* **517** (1999) 565

[3] D.N. Spergel et al., *Astrophys. J. Suppl.,* **148** (2003) 175

[4] M. Tegmark et al., *Phys. Rev. D,* **69** (2004) 103501

[5] G. Steigman and M.S. Turner, *Nucl. Phys. B,* **253** (1985) 375

[6] J.R. Primack, D. Seckel and B. Sadoulet, *Ann. Rev. Nucl. Part. Sci..* **38** (1988) 751

[7] G. Jungman, M. Kamionkowski and K. Griest, *Phys. Rept.,* **267** (1996) 195

[8] G. Servant et al., *New J. Phys.,* **4** (2002) 9

[9] D. Majumdar, *Phys. Rev. D,* **67** (2003) 095010

[10] J.D. Vergados, *Proc. dsu2006* (Madrid, 20-24 June 2006); hep-ph/0610017

[11] R.D. Peccei and H.R. Quinn, *Phys. Rev. Lett.,* **38** (1977) 1440

[12] S. Weinberg, *Phys. Rev. Lett.,* **40** (1978) 223

[13] F. Wilczek, *Phys. Rev. Lett.,* **40** (1978) 279

[14] G.G. Raffelt, *Phys. Rept.,* **198** (1990) 1

[15] M.S. Turner, *Phys. Rept.,* **197** (1990) 67

[16] S.J. Asztalos et al., *Phys. Rev. D.,* **69** (2004) 011101(R)

[17] L.D. Duffy, astro-ph/0603108

[18] G.G. Raffelt, hep-ph/061118; http://cast.mppmu.mpg.de/axion-training-2005/axion-training.php

[19] M.W.Goodman and E.Witten, *Phys. Rev. D*, **31** (1985) 3059

[20] P.F. Smith and J.D. Lewin, *Phys. Rept.,* **187** (1990) 203

[21] J. Carr et al., *Rep. Prog. Phys.,* **69** (2006) 2475

[22] M. Battaglia et al., *J. Phys. G,* **30** (2004) R217

[23] M. Nojiri et al., *J. High Energy Phys.,* **JHEP03** (2006) 063

[24] S. Desai et al., *Phys. Rev. D,* **70** (2004) 083523

[25] A. Kurylov and M. Kamionkowski, *Phys. Rev. D,* **69** (2004) 063503

[26] V.A. Bednyakov and H.V. Klapdor-Kleingrothaus, *Phys. Rev. D,* **63** (2001) 095005

[27] V. A. Bednyakov, F. Simkovic and I.V. Titkova, hep-ph/0412067

[28] R. Bernabei et al., *Riv. Nuovo Cim.,* **26N1** (2003) 1

[29] C. Savage, P. Gondolo and K. Freese, *Phys. Rev. D*, **70** (2004) 123513

[30] A.K. Drukier, K. Freese and D.N. Spergel, *Phys. Rev. D,* **33** (1986) 3495

[31] S. Cebrian et al., *Astropart. Phys.,* **14** (2001) 339; hep-ph/9912394

[32] D.N. Spergel, *Phys. Rev. D,* **37** (1988) 1353

[33] B. Morgan, A.M. Green and N.J.C. Spooner, *Phys. Rev., D* **71** (10) (2005) 103507

[35] authors in Proc. IDM96,98,2000,2002,2004,2006, ed. N.J.C. Spooner, World Sci. Press.

[36] S.P. Ahlen et al., *Phys. Lett. B*, **195** (1987) 603

[37] D.O. Caldwell et al., *Phys. Rev. Lett.,* **61** (1988) 510

[38] M. Beck et al., *Phys. Lett. B,* **141** (1994) 32

[39] L. Baudis et al., *Phys. Rev. D,* **63** (2001) 022001

[40] A. Morales et al., *Phys. Lett. B,* **532** (2002) 8; hep-ex/0110061

[41] A. Morales, *Nucl. Phys. B (Proc. Suppl.),* **138** (2005) 135

[42] S. Schonert et al., *Nucl. Phys. B (Proc. Suppl.),* **145** (2005) 242

[43] C.E. Aalseth et al., *Nucl. Phys. B (Proc. Suppl.),* **138** (2005) 217

[44] H. Klapdor-Kleingrothaus et al., *Eur. Phys. J. C,* **33** (2004) S962

[45] J.B. Birks, *The Theory and Practice of Scintillation Counting*, Pergamon Press, Oxford (1964)

[46] N.J.C. Spooner et al., *Phys. Lett. B,* **314** (1993) 430

[47] P.F. Smith et al., *Phys. Lett. B,* **379** (1996) 299

[48] R. Bernabei et al., *Phys. Lett. B,* **389** (1996) 757

[49] N.J.C. Spooner et al., *Phys. Lett. B*, **321** (1994) 156

[50] J.C. Barton and J.A. Edgington, *Nucl. Instrum. and Meth. in Phys. Res. A,* **443** (2000) 277

[51] B. Ahmed et al., *Astropart. Phys.,* **19** (2003) 691

[52] R. Bernabei et al., *Phys. Lett. B*, **424** (1998) 195

[53] R. Bernabei et al., *Phys. Lett. B*, **450** (1999) 448

[54] D. Akerib et al., *Phys. Rev. Lett.,* **93** (2004) 211301

[55] V. Sanglard et al., *Phys. Rev. D,* **71** (2005) 122002

[56] G.J. Alner et al. *Astropart. Phys.,* **23** (2005) 444

[57] C.J. Copi and L.M. Krauss, *Phys. Rev. D*, **67** (2003) 103507

[58] P. Ullio, M. Kamionkowski and P. Vogel, *J. High Energy Phys.,* **JHEP0107** (2001) 044

[59] C. Savage, P. Gondolo and K. Freese, *Phys. Rev. D,* **70** (2004) 123513

[60] R. Bernabei et al. *Eur. Phys. J. A,* **27** (2006) s1.57-s1.62

[61] R. Bernabei et al., *Nucl. Phys. B (Proc. Suppl.),* **138** (2005) 48

[62] S. Cebrian et al., *Nucl. Phys. B (Proc. Suppl.),* **114** (2003) 111

[63] S. Cebrian et al., *Nucl. Phys. B (Proc. Suppl.),* **138** (2005) 147

[64] H.S.Lee et al. *Proc. 9th Conf. CIPAN2006. AIP*, **870** (2006) 208; astro-ph/07040423

[65] N.J.C. Spooner et al., *Nucl. Instrum. and Meth. in Phys. Res. A*, **456** (2000) 272

[66] N.J.C. Spooner et al., *Phys. Lett. B,* **433** (1998) 150

[67] C. Angloher et al., *Astropart. Phys.,* **23** (2005) 325; astro-ph/0408006

[68] J. Ninkovic et al., *Nucl. Instrum. and Meth. in Phys. Res. A.,* **564** (2006) 567

[69] J. Hong et al. *Astropart. Phys.,* **16** (2001) 333

[70] G.J. Davies et al., *Phys. Lett. B*, **322** (1994) 159

[71] N.J.C. Sponer et al., Proc. IDM96, ed. N.J.C. Spooner, World Scientific (1997) 481

[72] Y. Shimizu et al., *Nucl. Instrum. and Meth. in Phys. Res. A,* **496** (2003) 347

[73] G.J. Alner et al., *Phys. Lett. B*, **616** (2005) 17

[74] P.F. Smith et al., *Phys. Lett. B*, **255** (1991) 454

[75] T.O. Niinikoski et al., *Nucl. Instrum. and Meth. in Phys. Res. A,* **559** (2006) 330

[76] N.J.C. Spooner et al., *Phys. Lett. B,* **273** (1991) 333

[77] T. Shutt et al., *Phys. Rev. Lett.,* **69** (1992) 3425

[78] D.S. Akerib et al., *Phys. Rev. D,* **73** (2006) 011102

[79] D.S. Akerib et al., *Phys. Rev. Lett.,* **96** (2006) 011302

[80] V. Sanglard et al., *Phys. Rev. D,* **71** (2005) 122002; astro-ph/0503265

[81] S. Cebrian et al., *Phys. Lett. B*, **563** (2003) 48

[82] G. Angloher et al. *Astropart. Phys.* **23** (2005) 325

[83] C. Arnaboldi et al., *Nucl. Instrum. and Meth. in Phys. Res. A,* **518** (2004) 775

[84] C. Arnaboldi et al., *Phys. Lett. B*, **584** (2004) 260

[85] C. Arnaboldi et al., *Astropart. Phys.,* **20** (2003) 91

[86] R. Bernabei et al., *Phys. Lett. B*, **436** (1998) 379

[87] R. Bernabei et al., *New J. Phys.,* **2** (2000) 15

[88] M.G. Boulay, A. Hime, and J. Lidgard, *Nucl. Phys. B (Proc. Suppl.),* **143** (2005) 486

[89] Y. Kim, *Phys. Atom Nucl.,* **69** (2006) 1970

[90] G.J. Alner et al., astro-ph/0701858

[91] G.J. Alner et al., astro-ph/0703362

[92] H.M. Araujo et al., *Astropart. Phys.,* **26** (2006) 140

[93] J. Angle et al., astro-ph/0609714; http://xenon.astro.columbia.edu/talks/APS2007

[94] M. Yamashita et al., *Astropart. Phys.*, **20** (2003) 79

[95] T.Doke. *Nucl. Instrum. and Meth. in Phys. Res. A,* **196** (1982) 87

[96] M.G. Boulay and A. Hime, *Astropart. Phys.,* **25** (2006) 179; astro-ph/0411358

[97] R. Brunetti et al., *New Astron. Rev.,* **49** (2005) 265

[98] P. Benetti et al., astro-ph/0701286

[99] A. Rubbia et al., *J. Phys. Conf. Ser.,* **39** (2006) 129

[100] M. Laffranchi et al., hep-ph/0702080

[101] D.N. McKinsey and K.J. Coakley, *Astropart. Phys.,* **22** (2005) 355; http://mckinseygroup.physics.yale.edu

[102] http://adsabs.havard.edu/abs/2007APS..APRB13005J

[103] E. Aprile et al., *Phys. Rev. Lett.,* **97** (2006) 081302

[104] P. Belli et al., *Il Nuovo Cimento C,* **19** (1996) 537

[105] P. Benetti et al., astro-ph/0603131

[106] V. Kudryavtsev, University of Sheffield, private communication (2006)

[107] T.A. Girard et al., *Phys. Lett. B,* **621** (2005) 233

[108] T. Morlat et al., *Nucl. Instrum. and Meth. in Phys. Res. A,* **560** (2006) 339

[109] M. Barnabe-Heider et al., *Phys. Lett. B,* **624** (2005) 186

[110] M. Barnabe-Heider et al., *Nucl. Instrum. and Meth. in Phys. Res. A,* **555** (2005) 184

[111] E. Moulin et al., *Nucl. Instrum. and Meth. in Phys. Res. A,* **548** (2005) 422

[112] L. Roszkowski et al., arXiv:0705.2012

[113] R. Trotta et al., *New Astron. Rev.,* **51** (2007) 316

[114] J. Ellis et al., *Phys. Rev. D,* **71** (2005) 095007

[115] E. Tziaferi et al., *Astropart. Phys.,* **27** (2007) 326

[116] M.J. Carson et al., *Astropart. Phys.,* **21** (2004) 667

[117] G. Alimonti et al., *Astropart. Phys.,* **16** (2002) 205; arXiv:physics/0702162

[118] J. Boger et al., *Nucl. Instrum. and Meth. in Phys. Res. A*, **449** (2000) 172

[119] P.K. Lightfoot et al., *Nucl. Instrum. and Meth. in Phys. Res. A*, **554** (2005) 266

[120] http://particleastro.brown.edu

[121] E. Aprile et al., *New Astron. Rev.,* **49** (2005) 289

[122] P. Majewski et al., *Proc. 7th Symp. on Dark Matter and Dark Energy in the Universe* (Marina del Rey, 22-24 Feb., 2006) tbp *Nucl. Phys. B. (Proc. Suppl.);* arXiv:astro-ph/0705.2117

[123] Giomataris et al., *Nucl. Phys. B (Proc. Suppl.),* **150** (2006) 208

[124] J. White et al., *Proc. 7th Symp. on Dark Matter and Dark Energy in the Universe* (Marina del Rey, 22-24 Feb., 2006) tbp *Nucl. Phys. B. (Proc. Suppl.)*

[125] D.S. Akerib et al., *Nucl. Instrum. and Meth. in Phys. Res. A,* **559** (2006) 411

[126] H. Kraus et al., *J. Phys. Conf. Ser.,* **39** (2006) 139

[127] G.J. Alner et al., *Nucl. Instrum. and Meth. in Phys. Res. A,* **535** (2004) 644

[128] G.J. Alner et al., *Nucl. Instrum. and Meth. in Phys. Res. A,* **555** (2005) 173

[129] D. Santos et al., *Proc. 3rd Symp. on Large TPCs (Paris,11-12 dec., 2006),* tbp *J. Phys. Conf. Ser.* (2007); arXiv:astro-ph/0703310

[130] A. Takada, et al., *Nucl. Instrum. and Meth. in Phys. Res. A,* **573** (2007) 195

[131] C.J. Martoff et al., *Nucl. Instrum. and Meth. in Phys. Res. A,* **555** (2005) 55

[132] A.M. Green and B. Morgan, *Astropart. Phys.,* **27** (2007) 142

[133] C.J. Copi, L.M. Krauss, *Phys. Rev. D*, **63** (2001) 043507

[134] C. J. Copi, J. Heo and L. M. Krauss, *Phys. Lett. B,* **461** (1999) 43

[135] N.J.C. Spooner et al., *Proc. IDM2006 (Rhodes, Greece 11-16[th] Sep., 2006)* tbp World Sci. (2007)

[136] P.K. Lightfoot et al., *Astropart. Phys. (2007)* Astropart. doi:10.1016/j.astropartphys.2007.02.003

[137] C.J. Martoff et al., *Nucl. Instrum. and Meth. in Phys. Res. A,* **526** (2004) 409

[138] A. Hitachi et al., *Proc. 3rd Symp. on Large TPCs (Paris,11-12 dec., 2006),* tbp *J. Phys. Conf. Ser.* (2007)

[139] C.J. Martoff et al., *Proc. 3rd Symp. on Large TPCs (Paris,11-12 dec., 2006),* tbp *J. Phys. Conf. Ser.* (2007)

[140] http://www.pppa.group.shef.ac.uk/cygnus2007

**Figures**

Figure 1: Summary of current spin-independent WIMP-nucleon limits (for references and details see text).

Figure 2: Plot of ionisation yield from recent XENON10 results showing the gamma background region (black dots) and the nuclear recoil selection region below (for details see ref. [93]).

Figure 3: Schematic design for the proposed miniCLEAN 100kg detector, a precursor to a potential ton-scale experiment (for details see text and ref. [101]).

Figure 4: Example 3D reconstruction of a 100 keV S recoil track obtained with the DRIFT II directional dark matter detector. Circles are indicative of the energy deposited along the track.

Figure 5: Example simulation of a 100 keV S recoil in 40 Torr $CS_2$ together with curves of dE/dx vs. energy for electronic and nuclear channels. Results indicate head-tail asymmetry is likely though experimental proof is needed that this can be observed in practice.

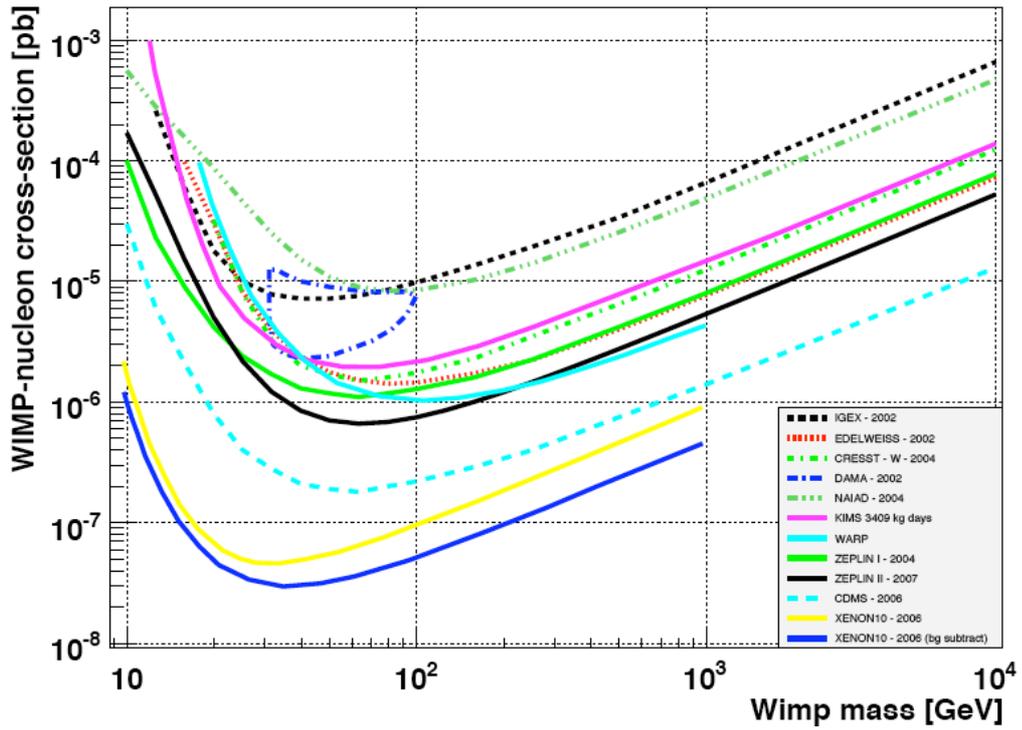

Figure 1: Summary of current spin-independent WIMP-nucleon limits (for references and details see text).

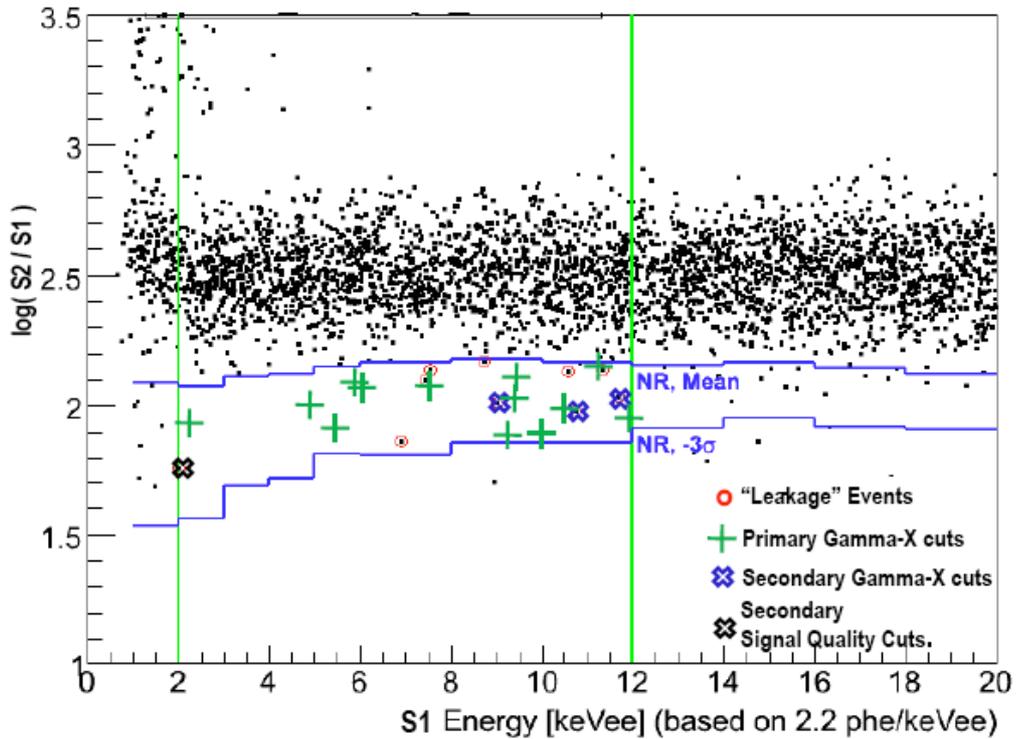

Figure 2: Plot of ionisation yield from recent XENON10 results showing the gamma background region (black dots) and the nuclear recoil selection region below (for details see ref. [93]).

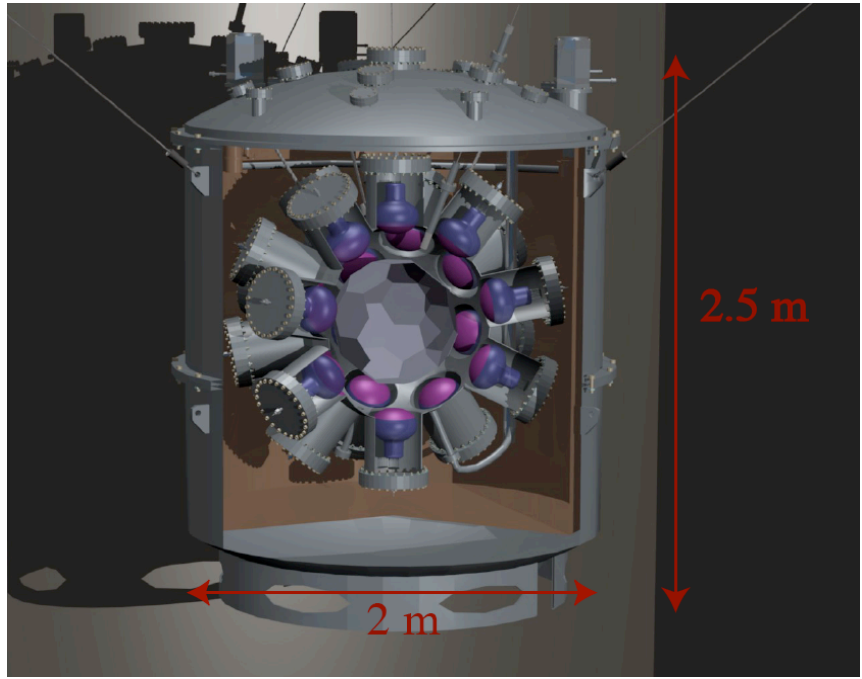

Figure 3: Schematic design for the proposed miniCLEAN 100kg detector, a precursor to a potential ton-scale experiment (for details see text and ref. [101]).

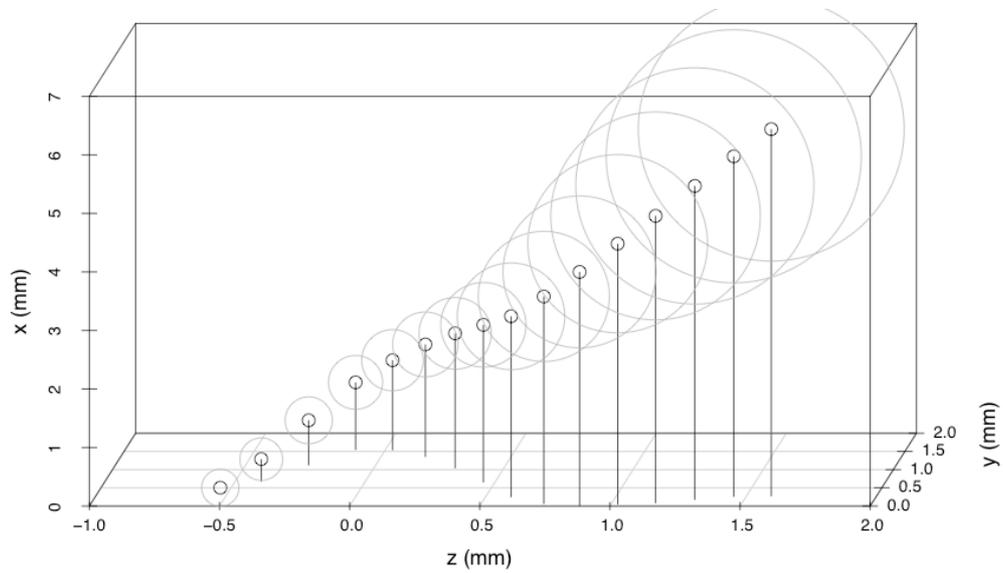

Figure 4: Example 3D reconstruction of a 100 keV S recoil track obtained with the DRIFT II directional dark matter detector. Circles are indicative of the energy deposited along the track.

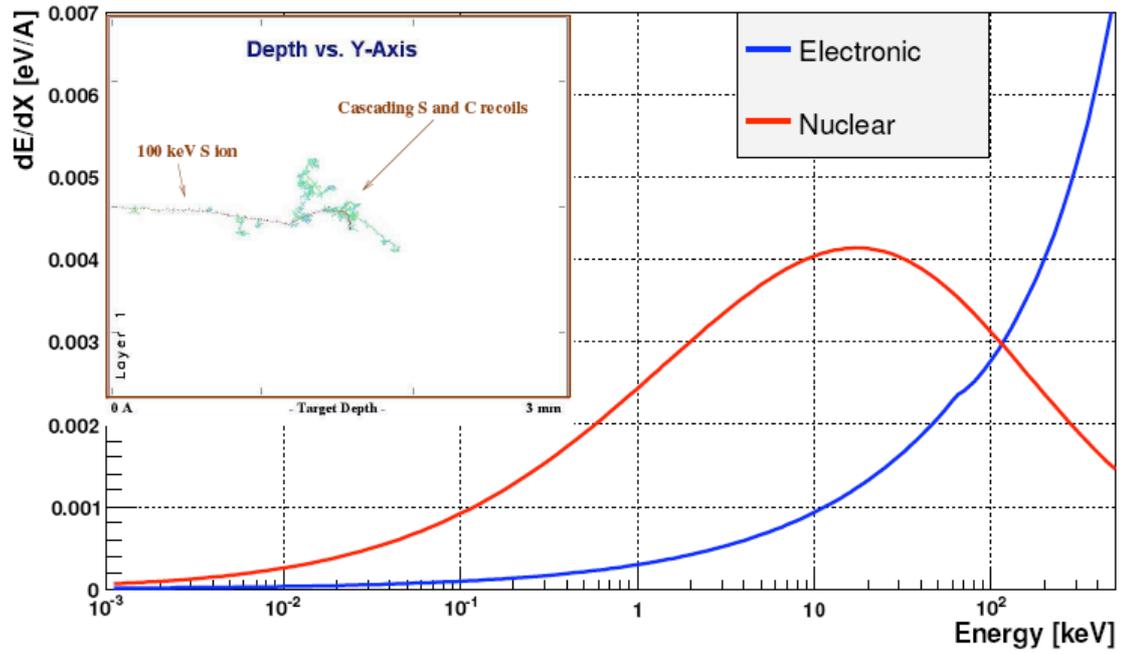

Figure 5: Example simulation of a 100 keV S recoil in 40 Torr $CS_2$ together with curves of dE/dx vs. energy for electronic and nuclear channels. Results indicate head-tail asymmetry is likely though experimental proof is needed that this can be observed in practice [135].